\documentclass[cits]{PoS}

\title{On the analytic continuation of the critical line}

\ShortTitle{On the analytic continuation of the critical line}

\author{Paolo Cea\\
        Dipartimento di Fisica dell'Universit\`a di Bari 
        and INFN - Sezione di Bari, \\I-70126 Bari, Italy\\
        E-mail: \email{paolo.cea@ba.infn.it}}

\author{Leonardo Cosmai\\
        INFN - Sezione di Bari, \\I-70126 Bari, Italy\\
        E-mail: \email{leonardo.cosmai@ba.infn.it}}

\author{Massimo D'Elia\\
        Dipartimento di Fisica dell'Universit\`a di Genova 
        and INFN - Sezione di Genova, \\I-16146 Genova, Italy\\
        E-mail: \email{massimo.delia@ge.infn.it}}

\author{Chiara Manneschi\\
        Dipartimento di Fisica dell'Universit\`a di Genova, 
        \\I-16146 Genova, Italy\\
        E-mail: \email{chiara.manneschi@ge.infn.it}}

\author{\speaker{Alessandro Papa}\\
        Dipartimento di Fisica, Universit\`a della Calabria, 
        and INFN - Gruppo Collegato di Cosenza, \\I-87036 Rende, Italy\\
        E-mail: \email{papa@cs.infn.it}}

\abstract{We perform a numerical study of the systematic effects involved in 
the determination of the critical line at real baryon chemical potential
by analytic continuation from results obtained at imaginary chemical 
potentials. We present results obtained in a theory free of the sign problem,
three-color QCD with finite isospin chemical potential, and comment on 
general features which could be relevant also to the continuation of the 
critical line in real QCD at finite baryon density.}

\FullConference{The XXVII International Symposium on Lattice Field Theory - LAT2009\\
		 July 26-31 2009\\
		 Peking University, Beijing, China}

\begin{document}

\section{Introduction}

The study of QCD at non-zero baryon density by numerical simulations
on a space-time lattice is plagued by the well-known sign problem:
the fermion determinant is complex and the Monte Carlo sampling becomes
unfeasible. One of the possibilities to circumvent this problem is to perform 
Monte Carlo numerical simulations for imaginary values of the baryon 
chemical potential, where the fermion determinant is real and 
the sign problem is absent, and to infer the behavior at real chemical 
potential by analytic continuation.
The idea of formulating a theory at imaginary $\mu$ was first suggested 
in Ref.~\cite{Alford:1998sd}, while the effectiveness of the method
of analytic continuation was pushed forward in Ref.~\cite{Lombardo:1999cz}.
Since then, the method has been extensively applied to  
QCD~\cite{muim,immu_dl,azcoiti,chen,defor06,Wu:2006su,sqgp,2im} and tested 
in QCD-like theories free of the sign problem~\cite{Hart:2000ef,giudice,
cea,cea1,conradi,Shinno:2009jw} and in spin models~\cite{potts3d,kt}. 
The state-of-the-art is the following: 

- the method is well-founded and works fine within the limitations posed 
by the presence of non-analyticities and by the periodicity of 
the theory with imaginary chemical potential~\cite{rw}; 

- the analytic continuation of physical observables is improved if ratios 
of polynomials (or Pad\'e approximants~\cite{Lombardo:2005ks}) are used as 
interpolating functions at imaginary chemical potential~\cite{cea,cea1}; 

- the analytic continuation of the (pseudo-)critical line on the
temperature -- chemical potential plane is well-justified, but a careful
test in two-color QCD~\cite{cea1} has cast some doubts on its reliability.

In particular, the numerical analysis in two-color QCD of Ref.~\cite{cea1} 
has shown that, while there is no doubt that an analytic function exists which 
interpolates numerical data for the pseudo-critical couplings for both
imaginary and real $\mu$ across $\mu=0$, determining this function
by an interpolation of data at imaginary $\mu$ could be misleading.
Indeed, in the case of polynomial interpolations, there is a clear indication
in two-color QCD that non-linear terms in $\mu^2$ play a relevant role at 
real $\mu$, but are less visible at imaginary $\mu$, thus calling for an 
accurate knowledge of the critical line there and, consequently, for
very precise numerical data. The above described scenario could well be 
peculiar to two-color QCD and strongly depend on the choice of parameters of 
Ref.~\cite{cea1}. Therefore, in this work we perform a systematic study of 
the analytic continuation of the critical line in another sign-free theory, 
SU(3) with a non-zero density of isospin. The dependence on the fermion mass
in two-color QCD is considered in a separate contribution~\cite{cea2}.
The aim of this study is to single out some general features of the 
analytic continuation of the critical line and to understand if
and to what extent they can apply also to the physically relevant case of QCD.

\begin{figure}[tbp]
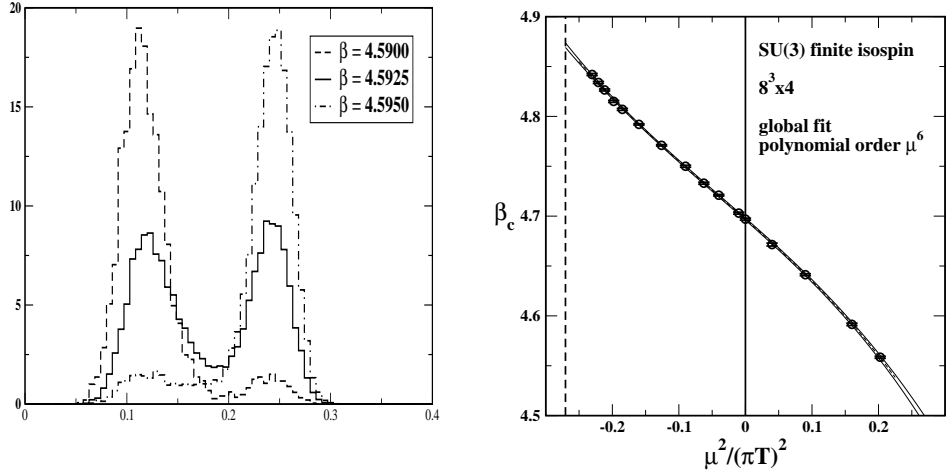

\vspace{0.2cm}
\centering
\includegraphics[height=6.5cm,width=0.38\textwidth]
{./figures/distribution_poly_su3.eps}
\hspace{0.5cm}
\includegraphics[width=0.4\textwidth]{./figures/fit_polin6_global_su3.eps}
\vspace{-0.2cm}
\caption[]{(Left) Distribution of the real part of the Polyakov loop 
in SU(3) with finite isospin density on a 8$^3\times 4$ lattice with 
$am$=0.1 at $\mu^2/(\pi T)^2$=0.16 and for three $\beta$ values around the 
transition. (Right) Critical couplings obtained in SU(3) with finite isospin 
density on a 8$^3\times 4$ lattice with $am$=0.1, together with a polynomial 
fit of order $\mu^6$ to {\em all} data.}
\label{su3_distributions}
\vspace{-0.1cm}
\end{figure}

\section{Analytic continuation of the critical line in 
three-color QCD at finite isospin chemical potential}

Three-color QCD with a finite density of isospin charge~\cite{ks} 
is a theory in which the chemical potential is $\mu$ for half of the 
fermion species and $-\mu$ for the other half. The partition function, which
is even in $\mu$ and depends only on $\mu^2$, can be 
written as follows:
\begin{equation}
Z(T,\mu) = \int \mathcal{D}U e^{-S_{G}} 
\det M [\mu]\det M [-\mu]\;,
\label{partfun}
\end{equation}
where the integration is over gauge link variables,
$S_G$ is the pure gauge action and $M$ the fermion matrix
(we adopt a standard staggered discretization).
This leads to a real and positive measure, because of the 
property $\det M [-\mu] = (\det M [\mu])^*$,
and therefore to a theory free of the sign problem. This theory is 
obviously closer to real QCD than two-color QCD, being yet unphysical,
since it implies a zero baryon density, while in Nature a non-zero
isospin density is always accompanied by a non-zero 
baryon density; moreover the isospin charge is not a conserved
number in the real world.
Nevertheless, for our purposes this theory is very convenient since it
provides us with another theoretical laboratory for the method of analytic 
continuation.

Similarly to SU(2) with finite baryon density, at imaginary values of the
chemical potential $\mu$ the theory exhibits RW-like transition lines, 
the first RW sector being given by the strip 
$-(0.5)^2 \lesssim \mu^2/(\pi T)^2 \leq 0$ 
(we refer to Ref.~\cite{cea3} for a detailed discussion of
the QCD phase diagram in presence of an imaginary isospin chemical 
potential). 

In our numerical analysis, we consider finite isospin SU(3) with $N_f=8$ 
degenerate staggered fermions of mass $am=0.1$ on a $8^3\times 4$ lattice. 
The critical line in the temperature -- chemical potential plane is a line of 
(strong) first order transitions, over all the investigated range of $\mu^2$ 
values, $-0.2304 \leq \mu^2/(\pi T)^2 \leq 0.2025$. This is one of the reasons 
for working on a small volume (tunneling between the different phases would 
have been sampled with much more difficulty on a larger volume) and clearly 
emerges from the distribution on the thermal equilibrium ensemble of the 
values of observables like the (real part of) the Polyakov loop, the chiral 
condensate, the plaquette across the transition (see, for example, 
Fig.~\ref{su3_distributions}(left)).
Typical statistics have been around 10K trajectories of 1 MD unit
for each run, growing up to 100K trajectories for 2-3 $\beta$ values
around $\beta_c(\mu^2)$, for each $\mu^2$, in order to correctly sample
the critical behavior at the transition. The critical $\beta(\mu^2)$ is 
determined as the point where the two peaks 
have equal height and, in all the cases considered, this point turned out
to be the same for all the adopted observables. 

\subsection{Results for the critical line at finite isospin}
\label{su3_results}

The general strategy is the following: after determining, for a set of 
$\mu^2$ values, the critical couplings $\beta_c (\mu^2)$, the critical line 
is guessed by interpolating the values of 
$\beta_c(\mu^2)$ for $\mu^2 \leq 0$ only. The validity of the interpolation
is evaluated by comparing its analytic continuation to the region 
$\mu^2 > 0$ with the direct determinations of the critical coupling in this 
region~\footnote{We refer to Ref.~\cite{cea3} for all the determinations of 
the critical couplings in the finite isospin SU(3) theory on a 8$^3\times 4$ 
lattice with fermion mass $am$=0.1 and for the parameters of all the fits
presented below.}.

We observe from the very beginning that data for $\beta_c(\mu^2)$ for both 
$\mu^2\leq 0$ and $\mu^2>0$ can be globally fitted by an analytic function 
(a polynomial of third order in $\mu^2/(\pi T)^2$ nicely works).
Fig.~\ref{su3_distributions}(right) shows how the fit compares with data.
The question is if there are interpolations of the critical couplings at 
$\mu^2 \leq 0$ only, that, when continued to $\mu^2 > 0$, agree with the 
critical couplings directly determined in the latter region.

\begin{figure}[tbp]
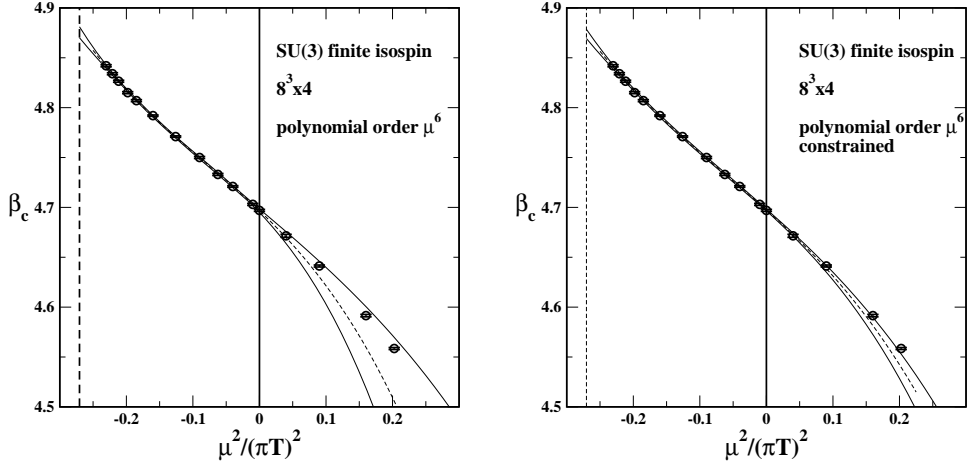

\vspace{0.2cm}
\centering
\includegraphics[width=0.4\textwidth]{./figures/fit_polin6_su3.eps}
\hspace{0.5cm}
\includegraphics[width=0.4\textwidth]{./figures/fit_polin_constrained_su3.eps}
\caption[]{(Left) Critical couplings obtained in SU(3) with finite isospin 
density on a 8$^3\times 4$ lattice with $am$=0.1, together with a polynomial 
fit of order $\mu^{6}$ to data with $\mu^2 \leq 0$. (Right) The same
with a polynomial of order $\mu^6$ with constrained quadratic term.}
\label{su3_crit_polin}
\end{figure}

We have tried several kind of interpolations of the critical couplings 
at $\mu^2 \leq 0$. 
At first, we have considered interpolations with polynomials up to
order $\mu^{10}$. We can see that data at $\mu^2 \leq 0$ are 
precise enough to be sensitive to terms beyond the order $\mu^2$; indeed,
a good $\chi^2$/d.o.f. is not achieved before including terms up to the 
order $\mu^6$, in agreement with the outcome of the global fit discussed
above. The extrapolation to $\mu^2 > 0$ for the polynomial of order 
$\mu^6$ is shown in Fig.~\ref{su3_crit_polin}(left); it agrees with direct 
determinations of $\beta_c(\mu^2)$, within the 95\% CL band.

Then, we have considered interpolations with ratios of polynomials of
order up to $\mu^{6}$. In all but one cases we got good fits to the
data at $\mu^2 \leq 0$, but only two extrapolations to $\mu^2 > 0$ compare
well with numerical data in that region: the ratio of a 4th to 6th order 
polynomial and the ratio of a 6th to 4th order polynomial~\cite{cea3}.
It is interesting to observe that the two interpolations which ``work'' 
have in common the number of parameters.

Both kinds of fits considered so far have evidenced that the role of
terms of order larger than $\mu^2$ cannot be neglected. Since the data
more sensitive to these terms are those farther from $\mu^2=0$, while data
closer to $\mu^2=0$ should ``feel'' only the $\mu^2$ term in a polynomial 
interpolation, we performed a fit with a polynomial of the form 
$a_0+a_1 \mu^2/(\pi T)^2$ in a small region $\mu^2 \lesssim 0$
and fixed the value of the parameter $a_1$ (see Ref.~\cite{cea3} for details).
Then, we kept $a_1$ fixed and repeated the fit on all available data at 
$\mu^2 \leq 0$ with a polynomial of the form $a_0+a_1 \mu^2/(\pi T)^2+a_2 
\mu^4/(\pi T)^4 +a_3 \mu^6/(\pi T)^6$. The resulting interpolation and its 
extrapolation to the region $\mu^2>0$ are shown in 
Fig.~\ref{su3_crit_polin}(right).
The comparison with critical couplings at $\mu^2>0$ is good and the 95\% CL
band is narrower than in the unconstrained $\mu^6$-polynomial fit 
(see Fig.~\ref{su3_crit_polin}(right)), meaning that this procedure
leads to increased predictivity for the method of analytic continuation.

\begin{figure}[tbp]
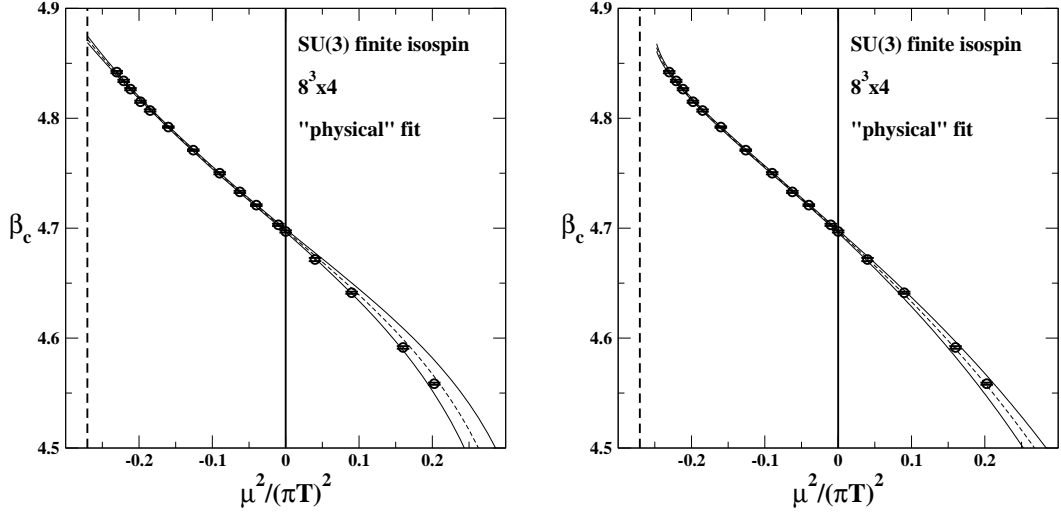

\vspace{0.3cm}
\centering
\includegraphics[width=0.44\textwidth]
{./figures/su3_phys_fit_2.eps}
\hspace{0.5cm}
\includegraphics[width=0.44\textwidth]
{./figures/su3_phys_fit_new_2.eps}
\caption[]{Values of $\beta_c(\mu^2)$ in SU(3) with finite isospin 
density on a 8$^3\times 4$ lattice with $am$=0.1, together with the 
fit to data with $\mu^2\leq 0$ according to the fit 
functions~(\ref{phys_fit})(left) and~(\ref{phys_crit_new})(right).} 
\label{su3_phys_fit}
\end{figure}

At last, we have attempted the fit strategy to write the interpolating 
function in {\em physical units} and to
deduce from it the functional dependence of $\beta_c$ on $\mu^2$, after
establishing a suitable correspondence between physical and lattice units.
The natural, dimensionless variables of our theory are $T/T_c(0)$,
where $T_c(0)$ is the critical temperature at zero chemical
potential, and $\mu/(\pi T)$. The question that we want to answer is 
if fitting directly the dependence of $T/T_c(0)$ on $\mu/(\pi T)$  
may lead to increased predictivity for analytic continuation. We shall name 
this kind of fits as ``physical'' fits.
While $\mu/(\pi T)$ is one of the dimensionless variables used in our 
simulations, $T/T_c(0)$ is not and must be deduced from the relation
$T=1/(N_t a(\beta))$, 
where $N_t$ is the number of lattice sites in the temporal direction
and $a(\beta)$ is the lattice spacing at a given $\beta$~\footnote{
Strictly speaking the lattice spacing depends also on the bare quark 
mass, which in our runs slightly changes as we change $\beta$ since we fix $a m$.
However in the following evaluation, which is only based on the perturbative
2-loop $\beta$-function, we shall neglect such dependence.
}. 
Since our determinations for $\beta_c$ range between $\simeq$ 4.5585 and 
$\simeq$ 4.842, it can make sense to use for $a(\beta)$ the perturbative 
2-loop expression with $N_c=3$ and $N_f=8$.

We have tried several different fitting functions and report two cases
which work particularly well. The first is given by the following
3-parameter function:
\begin{equation}
\left[\frac{T_c(\mu)}{T_c(0)}\right]^2=\frac{1+B\mu^2/(\pi T_c(\mu))^2}
{1+A\mu^4/(\pi T_c(\mu))^4}\; 
\label{phys_crit}
\end{equation}
leading to the following implicit relation between $\beta_c$ and 
$\mu^2$:
\begin{equation}
a(\beta_c(\mu^2))^2\biggr|_{\rm 2-loop} 
= a(\beta_c(0))^2\biggr|_{\rm 2-loop}
\; \frac{1+A\mu^4/(\pi T_c(\mu))^4}{1+B\mu^2/(\pi T_c(\mu))^2}\;.
\label{phys_fit}
\end{equation}
In Fig.~\ref{su3_phys_fit}(left) we compare to data the fit with the 
function~(\ref{phys_fit}): one can see that the extrapolation to the
region $\mu^2>0$ behaves very well.
The values of the fit parameters are $\beta_c(0) = 4.6977(13)$,
$A = -3.25(26)$ and $B = -2.62(12)$, with $\chi^2$/d.o.f.=1.33. 

As an alternative function for the shape of the critical line, we have 
tried also the following
\begin{equation}
\frac{T_c(\mu)}{T_c(0)}=
\left\{
\begin{array}{ll}
A+(1-A)\left[\cos\left(\frac{\mu}{T}\right)\right]^B\;, & \mu^2\leq 0 \\
A+(1-A)\left[\cosh\left(\frac{\mu}{T}\right)\right]^B\;, 
& \mu^2 > 0 \;, \\
\end{array}
\right.
\label{phys_crit_new}
\end{equation}
which explicitly encodes the expected periodicity of the partition
function for imaginary $\mu$.  
The fit to data at imaginary $\mu$ is very good and its extrapolation
to the real chemical potential side compares impressively well with data
(see Fig.~\ref{su3_phys_fit})(right). The resulting fit parameters are
$\beta_c(0) = 4.6969(12)$, $A = 1.508(15)$ and $B= 0.560(32)$, with
$\chi^2$/d.o.f.=0.39. This function is a good candidate to 
parameterize the critical line for small values of $\mu/T$.

In both cases, Eq.~(\ref{phys_fit}) and Eq.(\ref{phys_crit_new}), the 
``physical'' fit worked very well and with a reduced number of parameters
with respect to our previous fits, leading to increased predictivity and 
consistency with data at real chemical potentials. One can easily 
check that the adopted functions are not appropriate for a continuation 
of the critical line down to the $T = 0$ axis, but this is not the aim 
of our study, since such extrapolation would be questionable anyway. 

\section{Conclusions}

In this work we have presented results concerning the analytic continuation 
of the critical line in QCD with a finite density of isospin charge.
We have detected some features and developed some strategies,
which could apply and be useful for real QCD at finite baryon density. Let 
us briefly summarize them.

- Non-linear terms in the dependence of the pseudocritical 
coupling $\beta_c$ on $\mu^2$ in general cannot be neglected.
A polynomial of order $\mu^6$ seems to be sufficient in all explored cases. 

- The coefficients of the linear and non-linear terms in $\mu^2$ in
a Taylor expansion of $\beta_c(\mu^2)$ are all negative. That often implies
subtle cancellations of non-linear terms at imaginary chemical 
potentials ($\mu^2 < 0$) in the region available for analytic
continuation (first RW sector). The detection of such terms,
from simulations at $\mu^2 < 0$ only, may be difficult and requires an
extremely high accuracy.
As a matter of fact, the simple use of a sixth order polynomial to fit 
data at imaginary $\mu$ leads to poor predictivity, which is slightly 
improved if ratio of polynomials are used instead.

- An increased predictivity is achieved if the linear term in $\mu^2$ is 
fixed from data at small values of $\mu^2$ only. 

- We have proposed a new, alternative ansatz to parameterize the
critical line directly in physical units in the $T,\mu$ plane 
(instead than in the $\beta,\mu$ plane) and given two explicit realizations.
This ``physical'' ansatz provides a very good description of the critical 
line, moreover with a reduced number of parameters, and leads to an increased 
predictivity, comparable to that achieved by the ``constrained'' fit.

\end{document}